\documentclass[a4paper]{jpconf}
\usepackage{graphicx}

\begin{document}
\title{A Silicon MEMS EM vibration energy harvester}
\author{Y. Yang, U. Radhakrishna, D. Ward, A. P. Chandrakasan \newline and J. H. Lang}
\address{ MIT, Cambridge, MA, 02139, USA.}
\ead{markyang@mit.edu}
\begin{abstract}
This paper presents an optimized silicon-MEMS electromagnetic vibration energy harvester suitable for applications such as machine health monitoring. The harvester comprises a DRIE-etched silicon suspension, and pick-and-place N42 NdBFe magnets and copper coils, housed in a 3D-printed package. The harvester is designed to operate near 50 Hz with 0.5-1~$g$ vibrations using a long-stroke suspension. Multi-domain harvester optimization results in an open-circuit voltage of 1.7 V, a matched-load power output of 2.2 mW, and a matched-load power-output density of 1.23~mW/cm$^3$ at 1.1 $g$ with a resonance frequency of $76.3\ Hz$.
\end{abstract}

\section{Introduction}
This paper presents a vibration energy harvester, employing a Lorentz-force energy converter, that is suitable for powering miniaturized autonomous IoT sensors \cite{Arnold}\cite{Y.Tan}. The harvester comprises a DRIE-etched silicon suspension, and pick-and-place N42 NdBFe magnets and copper coils, all enclosed in a 3D-printed plastic package. The harvester has an active volume of 1.79~cm$^3$, and an output power $P_{\rm Out}$ of 2.2~mW at $1.1\ g$  and 76 Hz under matched load, yielding a power density (PD) of 1.23~mW/cm$^3$ and a normalized power density (NPD) of 1.02~mW/cm$^3/g^2$, the highest reported PD and NPD among silicon-based MEMS harvesters reported to date \cite{Y.Tan}. The high $P_{\rm Out}$ follows the use of a four-bar-linkage suspension that lowers beam stress compared to our earlier accordion suspension \cite{Abraham}, enabling mm-range strokes and hence mW-level $P_{\rm Out}$. The key contributions here are: (i) large-stroke (2 mm) silicon suspensions with stress analysis, (ii) harvester implementation yielding $P_{\rm Out}=2.2$ mW, and (iii) optimized design guidelines and scaling to further reduce harvester size while preserving $P_{\rm Out}$.

\section{Design and Optimization}
As described below, harvester design is based on optimizations over mechanical, magnetic and electrical domains. To begin, a mechanical optimization of the harvester spring-mass-damper system is executed following \cite {Abraham}. To do so, the mechanical power $P_{\rm M}$ converted by the harvester is expressed at resonance in sinusoidal steady state in the absence of mechanical loss as
\begin{equation} 
P_{\rm M} = B \omega^2 X^2 / 2 = \omega X M A / 2 = \rho \omega X A (L_1 - 2S) L_2 L_3 / 2 
\label{OPT1}
\end{equation}
where damping coefficient $B$ is a proxy for energy conversion through the Lorentz-force energy converter, $\omega$ is the resonance frequency, $X$ is the stroke amplitude, $M$ is the proof-mass mass, $A$ is vibration acceleration, $\rho$ is the density of the proof mass, $L_1$, $L_2$ and $L_3$ are the dimensions of the harvester with $L_1$ in the stroke direction, and $S \ge X$ is the single-sided space within $L_1$ allocated for stroke; the assumption of negligible mechanical loss is supported by experimental observation. Following \cite{Abraham}, $P_{\rm M}$ is maximized when $X=S=L_1/4$, yielding
\begin{equation} 
P_{\rm M,Max} =  \rho \omega S A (L_1 - 2S) L_2 L_3 / 2 =  \rho \omega A L_1^2 L_2 L_3 / 16
\quad . 
\label{OPT2}
\end{equation}
Thus, $L_1$, is allocated equally to the mass and bi-directional stroke $2X$. This assumes the suspension/springs permit (nearly) all of $S$ to be used for $X$, which was true for the accordion suspension \cite{Abraham}. Finally, following (\ref{OPT2}), the longest harvester dimension is used for the stroke direction $L_1$ for maximum $P_{\rm M}$. Note that to achieve the maximized power in (\ref{OPT2}), the Lorentz-force energy converter must provide the damping $B$ required to satisfy (\ref{OPT1}) and $X=L_1/4$.

\begin{figure}[tb]\hspace{-0.4 cm}
\includegraphics[width=39pc, clip=true, trim=0mm 0mm 0mm 0mm]{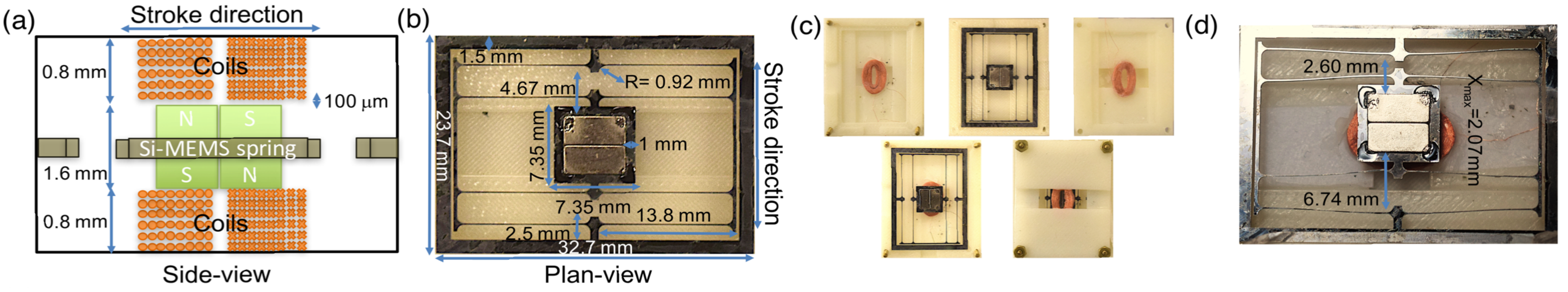}
\caption{ (a) Cross-sectional side view of the harvester. (b) Top view of the suspension and magnets glued together. (c) Individual 3D-printed packaging parts to be combined. (d) Harvester at its maximum deflection showing the bending of the beams.
}\label{device}
\vspace{-1.2mm}
\vspace{-1.2mm}
\end{figure}

To achieve $P_{\rm Out}$ in the mW range, the stroke $X$ must be in the mm range. Our previous accordion suspension \cite{Abraham} did not yield large $X$ as the springs fractured at $X=0.6$~mm due to high stress caused by the side-bar used to raise the resonant frequencies of the higher-order vibration modes. A four-bar-linkage suspension is chosen here to overcome the spring fracture at large strokes while raising the resonance frequencies of the higher-order modes. Figures \ref{device}(a)-(b) show a cross-sectional side-view of the fabricated harvester and a top-view of the silicon four-bar-linkage suspension.
%
\vspace{-1.2mm}
The four-bar linkage is not as space efficient as the accordion suspension as it requires $X\le S/2$. $P_{\rm M}$ in (\ref{OPT1}) is then maximized with $X=S/2=L_1/8$, yielding
\begin{equation} 
P_{\rm M,Max} =  \rho \omega S A (L_1 - 2S) L_2 L_3 / 4 =  \rho \omega A L_1^2 L_2 L_3 / 32
\quad .
\label{OPT3}
\end{equation}
However, the springs in the new suspension should experience lower stress than in \cite{Abraham} because they are not torqued by the side bar. This is evident from the spring bending at maximum stroke shown in Figure \ref{device}(d). Second, the spring widths are tapered via narrowing towards their middle to create a more uniform stress profile. Third, the joints are designed to have symmetric fillets with radius of five spring widths to further reduce stress. Fourth, the spring length is increased to reduce strain. These precautions yield the desired stroke of $X=S/2=2$ mm as shown in Figure \ref{device}(d), the largest stroke reported in a Si-MEMS harvester suspension \cite{Y.Tan}.

\begin{figure}[tb]
\vspace{-1.9 cm}
\hspace{0.9 cm}\includegraphics[width=34pc, clip=true, trim=0mm 0mm 0mm 0mm]{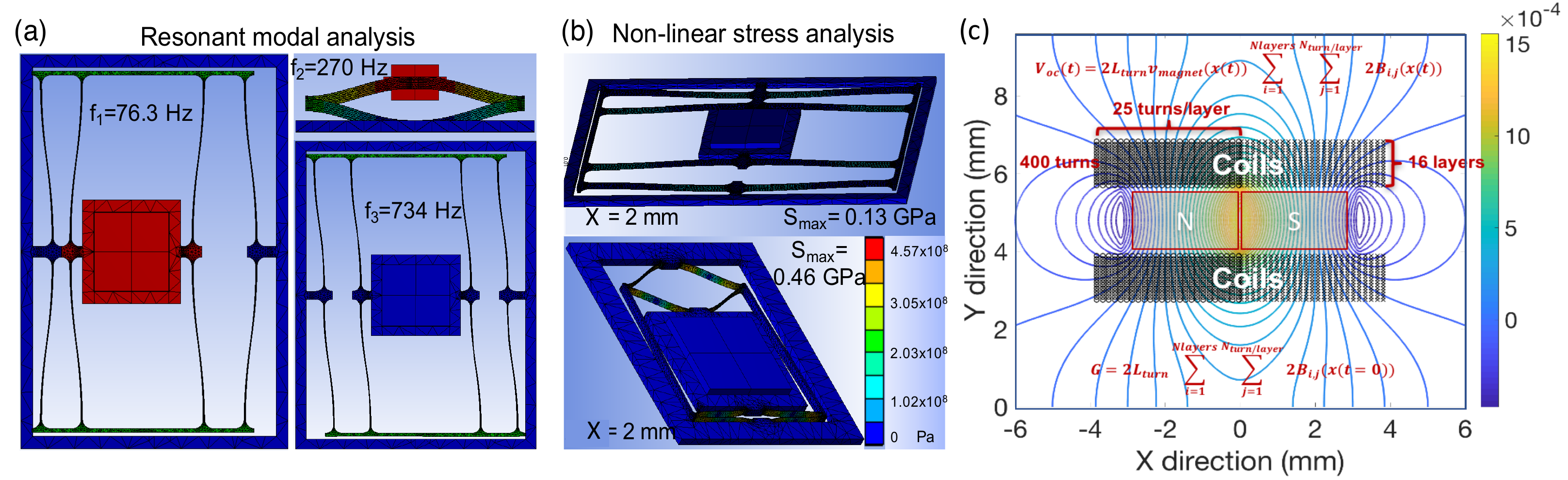}\vspace{-0.3 cm}
\caption{\label{resonance} (a) Resonant modal analysis of the 4-bar-linkage suspension showing good modal separation between higher resonant modes and the desired fundamental mode. (b) Stress analysis showing that the suspension under maximum stroke has three times lower stress than does the previous accordion spring suspension \cite{Abraham}. (c) 2D magnetic simulation to compute $G$.
}
\vspace{-1.2mm}
\vspace{-1.2mm}
\end{figure}

The suspension is dimensioned to achieve a
%
near-50-Hz resonance subject to the constraints that the spring width exceeds 25 $\mu$m as limited by etch resolution, and beam stress is less than 130~MPa. Figure \ref{resonance}(a) shows the resonant modal analysis for the optimized suspension highlighting a fundamental resonance translational mode $f_{\rm Res}$ close to 50  Hz and a 5-fold separation between $f_{\rm Res}$ and higher-order resonance frequencies.
Solidworks simulations show the lowest resonant modes are: translational along $L_1$ at $76\ Hz$; translational along the $L_3$ (magnetic pole) direction at 270 Hz; and rotational about $L_3$ at 575 Hz. The stress analysis in Figure \ref{resonance}(b) also shows that the maximum beam stress at full stroke is 130 MPa compared to 460 MPa in the accordion suspension. Thus, the mechanical optimization achieves optimum resource allocation for mass and stroke while the suspension exhibits the desired $f_{\rm Res}$ and $X$. 

Two anti-parallel N42 NdBFe permanent magnets are the magnetic flux source through which $B$ is implemented, and the proof mass because they offer higher mass density than loose windings. Figure \ref{device}(a) shows the magnetic energy converter cross section with moving magnets, and two stationary coils on the harvester frame. The design parameters of magnet height $T$, coil height $\Delta$, and airgap between magnet and coil $\delta$ are optimized for maximum time-average electrical output power $P_{\rm E}$. Assuming the magnetic thickness of $T+2\Delta+2\delta$ is small compared to the magnet widths and lengths, the maximum magnetic flux density, and hence the maximum induced coil voltage $V$ at a given $\omega$ and $X$, the coil resistance $R_{\rm Coil}$, and the maximum output power $P_{\rm E} = V^2/8R_{\rm Coil}$ into a matched load, are given by
\begin{equation} 
V \propto NT/(T+2\Delta+2\delta)\quad ;\quad
R_{\rm coil} \propto N^2/\Delta\quad ;\quad
P_{\rm E} \propto \Delta T^2/(T+2\Delta+2\delta)^2
\end{equation}
with $N$ coil turns. For a given magnet size and hence $T$, $P_{\rm E}$ is maximized with the smallest permissible $\delta$, and $\Delta = T/2 + \delta$. A harvester so optimized produces a maximum induced coil voltage in sinusoidal steady state given by Faraday's Law according to
\begin{equation} 
V = d\phi_B/dt= B L_2 \omega X = G \omega X \quad .
\end{equation}
2D-magnetic simulation computes the flux distribution and output voltage as described in \cite{Abraham}; a magnetization of 1.3 T is chosen for the permanent magnets. The magnetic flux distribution of $B_x$ and $B_y$ is shown for zero stroke in Figure \ref{resonance}(c). Together with the velocity $u(t)$, the distribution is used to compute the time-dependent voltage across each coil turn according to
\begin{equation} 
 V_{\rm Turn}(t)= \omega L_3 (A(x_1(t))-A(x_2(t)) = u(t) L_3 (B_y(x_1(t))-B_y(x_2(t)) = u(t) G(t) \quad .
\end{equation}
Finally, an optimized coil configuration maximizes the mechanical-to-electrical transduction coefficient $G$ while minimizing the coil resistance $R_{\rm Coil}$. The total number of layers $N_{\rm Layers}$ of 50-$\mu$m-diameter copper-wire coils within $T$, and the number of turns $N_{\rm Turns}$ in each such layer, are determined based on this optimization. While thick coils reduce $R_{\rm Coil}$, they also result in fewer $N_{\rm Layers}$ within a given $T$ and hence a lower $G$. The electrical optimization considers this trade-off using the metric $G^2/R_{\rm Coil}$ as the optimization goal to maximize power output.

\section{Fabrication and Packaging}
\begin{figure}[bt]
\vspace{-0.5 cm}\includegraphics[width=38pc, clip=true, trim=0mm 0mm 0mm 0mm]{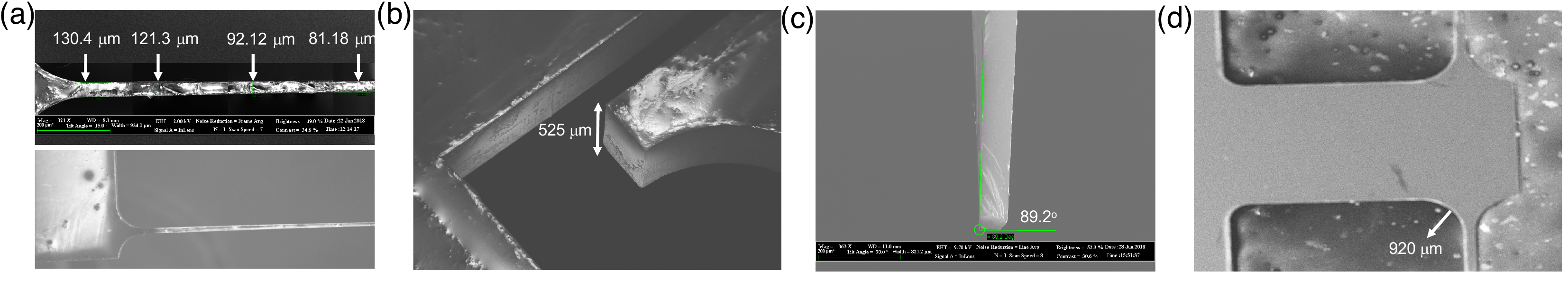}
\caption{\label{sem} SEM images. (a) Tapered spring profile showing the spring width variation. (b) Angled view of a joint. (c) Side view of an inner suspension wall showing the etch profile. (d) An image of the anchor with the fillet profile.}
\vspace{-1.2mm}
\end{figure}

Silicon-spring suspensions are fabricated using a single deep-reactive-ion etch through a 525-$\mu$m-thick wafer, with the mask haloed \cite{Abraham}-\cite{blowout} using 40-$\mu$m-wide trenches to reduce etch loading, resulting in essentially vertically-etched side walls as shown in Figure \ref{sem}. Additionally, the mask is biased to accommodate a 5-$\mu$m blow-out per side wall. This fabrication process allows a minimum feature size of 25 $\mu$m, smaller than the minimum spring width of 30 $\mu$m used in the suspension. SEM images in Figure \ref{sem} show the tapering of the beams along their length, and vertical smooth surfaces after etching that prevent damping. The magnets are glued inside the Si-wrapper linked to the suspension using a pick-and-place process with a 3D-printed platform. The spring-mass system is housed within another plastic assembly shown in Figure \ref{device}(c) along with parts that hold the two coils that are wound from 44 AWG wire on a Mandrell using a lathe. Each coil has 400 turns resulting in an individual $R_{\rm Coil}=123\ \Omega$. The full assembly with the plan- and side-views is shown in Figure \ref{device}.

\section{Experimental Performance} 
\begin{figure}[tb]
\hspace{-0.4 cm}\includegraphics[width=40pc, clip=true, trim=0mm 0.3mm 0mm 1mm]{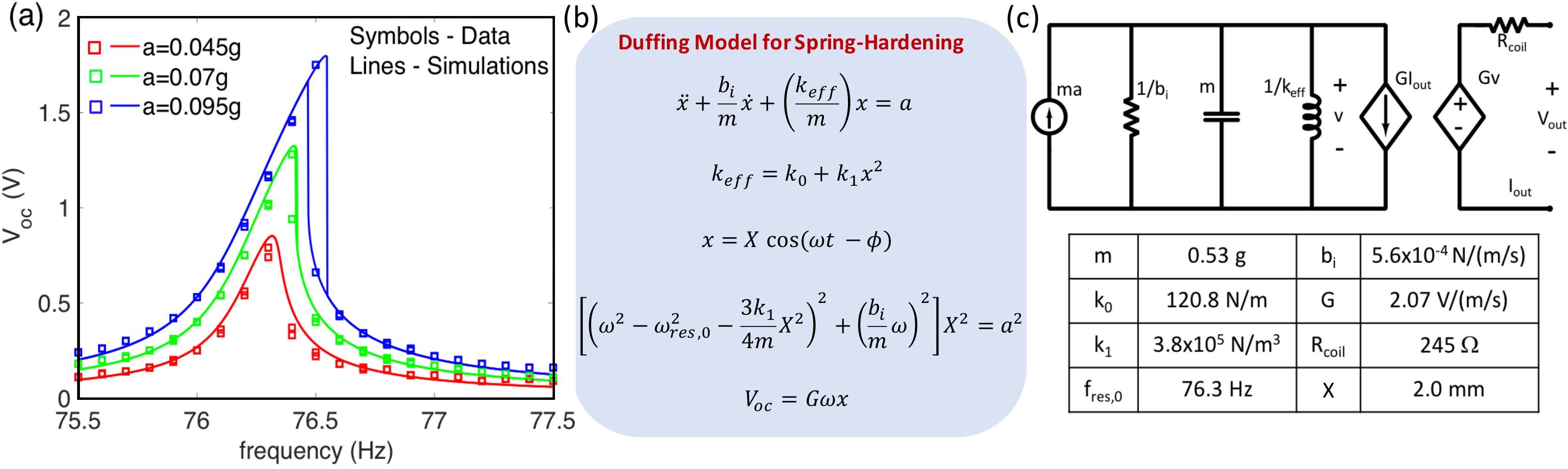}
\caption{\label{opencircuit} (a) Measured open-circuit voltage compared against simulations for different $\omega$ and $g$. (b) Harvester dynamic model with the Duffing spring hardening. (c) Harvester equivalent circuit model with extracted mechanical and electromagnetic parameters.
\vspace{-1.2mm}
\vspace{-1.2mm}
\vspace{-1.2mm}
}
\end{figure}

Harvester parameters are extracted from the measured open-circuit voltage $V_{\rm OC}=G \omega X$ as a function of $\omega$ and $g$; see Figure \ref{opencircuit}(a). Various optimizations involved in harvester design result in the high $V_{OC}=1.75$ V. Figure \ref{opencircuit}(a) shows that the suspension springs harden at higher $g$, exhibiting two voltage branches together with a slight increase in $\omega_{\rm Res}=2\pi f_{\rm Res}$ with $g$. The Duffing dynamics \cite{duffing} used to fit the data and extract $X$ are shown in Figure \ref{opencircuit}(b) while the harvester equivalent circuit model and its parameters are listed in Figure \ref{opencircuit}(c). The parameters are extracted from the two lower-$g$ data and the model fits well against the data for $g=0.095$.

\begin{figure}[bt]
\includegraphics[width=38pc, clip=true, trim=0mm 0mm 0mm 0mm]{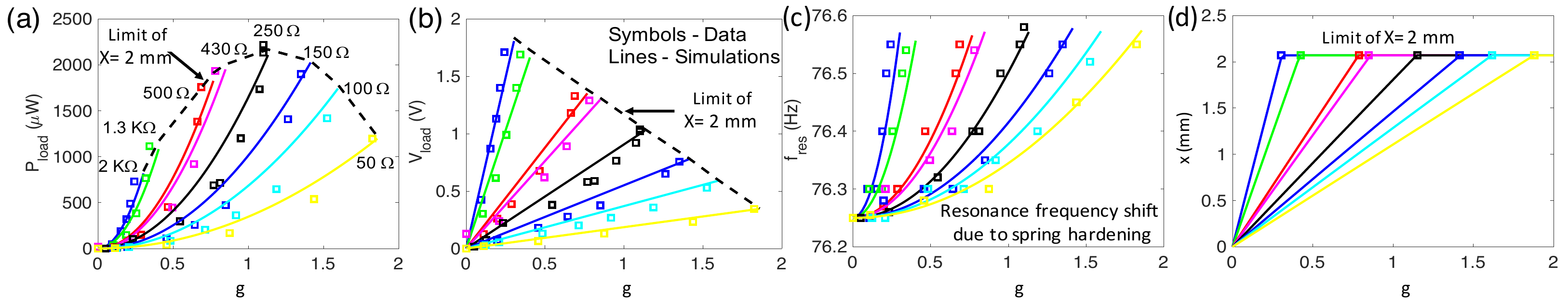}
\caption{\label{power} Measured and simulated (a) $P_{\rm Out}$, (b) $V_{\rm Load}$, (c) $f_{\rm Res}$ and (d) $x_{\rm Max}$ as functions of $g$ for different $R_{\rm Load}$ at resonance. 

}
\end{figure}

The measured loaded output power $P_{\rm Out}$, output voltage $V_{\rm Load}$, resonance frequency $f_{\rm Res}=\omega_{\rm Res}/2\pi$, and stroke X are plotted against $g$ for different loads $R_{\rm Load}$. The measurements, all performed at resonance, are compared against model simulations in Figure~\ref{power} showing a good match. The maximum $P_{\rm Out}$ of 2.2 mW is obtained with the matched load of $R_{\rm Load}=R_{\rm Coil}=245\ \Omega$ at 1.1 $g$, the highest reported value to date for Si-based MEMS harvesters, at least at sub-kHz frequencies. The device is currently stroke-limited with the side-bar hitting the external frame. If $R_{\rm Load}>R_{\rm Coil}$, reduced electrical damping causes the stroke to reach its maximum of $X = S/2$ at a lower $g$ with a higher $V_{\rm Load}$. The opposite holds for $R_{\rm Load}<R_{\rm Coil}$ with lower $V_{\rm Load}$ at $X=S/2$ at higher $g$. The harvester has a volume of 1.79 cm$^3$ which results in performance metrics of: PD = 1.23 mW/cm$^3$, NPD = 1.02 mWcm$^{-3}g^{-2}$, and PD$/g/f_{\rm Res}=15 \mu$Wcm$^{-3}g^{-1}$s$^{-1}$ \cite{mitcheson}. The PD is highest among reported EM harvesters \cite{Y.Tan}, and the NPD is highest among MEMS harvesters (Table 3 of \cite{Y.Tan}). The NPD ranks second against non-Si, non-MEMS harvesters (behind \cite{beeby} in Table I of \cite{Y.Tan}). 

\begin{figure}[tb]
\hspace{-0.2 cm}\includegraphics[width=39pc, clip=true, trim=0mm 1mm 0mm 1mm]{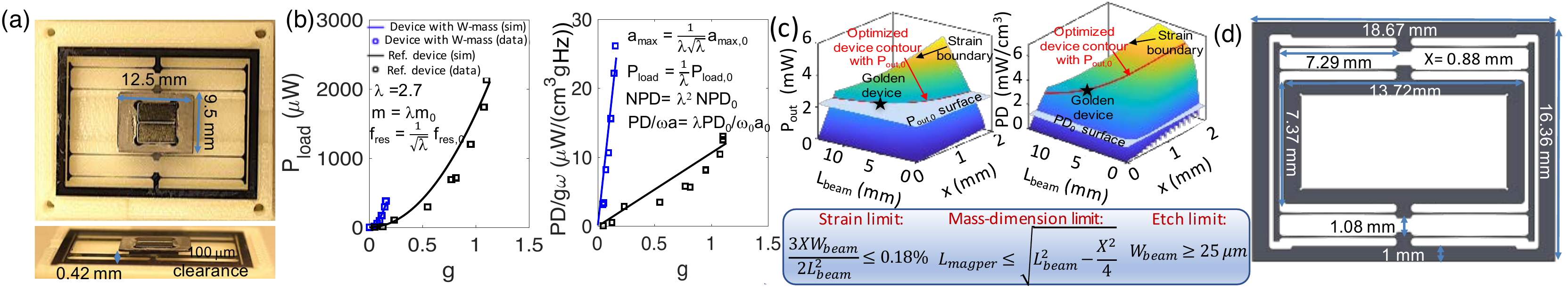}
\caption{\label{augmass} (a) Proof mass augmented with a Tungsten magnet wrapper. (b) Resulting lower-$g$ operation yielding increased NPD and PD$/g/f$; also scaling rules. (c) Mechanical optimization to lower the overall dimensions by making magnet and spring dimensions equal allows higher PD for the same $P_{\rm Out}$. (d) Cross-section of the optimized golden device.
\vspace{-1.2mm}
\vspace{-1.2mm}
\vspace{-1.2mm}
}
\end{figure}

A Tungsten wrapper is added around the magnet/proof-mass to increase the mass by 2.7 fold, reducing the acceleration to achieve $X=2.07$ mm as shown in Figure \ref{augmass}. The measured $P_{\rm Out}$ and PD$/g/f_{\rm Res}$ compared against scaling-law simulations show lower PD = 212 $\mu$W/cm$^3$, but higher NPD = 8951 $\mu$Wcm$^{-3}g^{-2}$ and PD$/g/f_{\rm Res}=29\ \mu$Wcm$^{-3}g^{-1}$s$^{-1}$, the highest among EM-harvesters reported in \cite{Y.Tan} if swept volume is considered. Shortening the spring and $X$ to reduce the harvester footprint, and increasing magnet width in the unused $L_2$ direction, yield a ``golden-device'' projected to provide similar $P_{\rm Out}=2.2$ mW at reduced $X=0.88$ mm and a volume of 0.59 cm$^3$, but at the same $f_{\rm Res}$ and operating $g$ as shown. This harvester has improved PD$=3729\ \mu$W/cm$^3$, NPD $=3082\ \mu$Wcm$^{-3}g^{-2}$, and PD/$g/f_{\rm Res}=44.5\ \mu$Wcm$^{-3}g^{-1}s^{-1}$.
\vspace{-1.2mm}
\vspace{-1.2mm}

\section{ Summary \& Conclusions}
This paper presents a compact MEMS electromagnetic vibration energy harvester designed for near 50-Hz operation. Optimization, and design guidelines are provided together with fabrication details. The harvester has a volume of 1.79~cm$^3$, and demonstrates an open-circuit voltage of 1.75 V and a matched-load power of 2.2 mW at 1.1 g near 76 Hz. This demonstrates that small vibration energy harvesters can provide the power required by autonomous machine health monitoring for industrial IoT and other remote sensing applications.
\vspace{-1.2mm}
\section*{Acknowledgments}
The authors wish to thank Analog Devices Inc.\ for collaboration and project support.
\vspace{-1.2mm}


\end{document}